\documentclass[aip,apl,reprint]{revtex4-1}

\pdfoutput=1
\usepackage{graphicx}
\usepackage{amsmath}

\begin{document}


\title{Microscopic magnetic structuring of a spin-wave waveguide by ion implantation in a Ni$_\text{81}$Fe$_\text{19}$ layer} 



\author{Bj\"orn~Obry}
\author{Thomas~Meyer}
\author{Philipp~Pirro}
\affiliation{Fachbereich Physik and Forschungszentrum OPTIMAS, Technische Universit\"at Kaiserslautern, D-67663 Kaiserslautern, Germany}

\author{Thomas~Br\"acher}
\affiliation{Fachbereich Physik and Forschungszentrum OPTIMAS, Technische Universit\"at Kaiserslautern, D-67663 Kaiserslautern, Germany}
\affiliation{Graduate School Materials Science in Mainz, D-67663 Kaiserslautern, Germany}

\author{Bert L\"{a}gel}
\affiliation{Nano Structuring Center, Technische Universit\"{a}t Kaiserslautern, D-67663 Kaiserslautern, Germany}

\author{Julia~Osten}
\author{Thomas~Strache}
\author{J\"urgen~Fassbender}
\affiliation{Institut f\"ur Ionenstrahlphysik und Materialforschung, Helmholtz-Zentrum Dresden-Rossendorf, D-01328 Dresden, Germany, and Technische Universit\"{a}t Dresden, D-01062 Dresden, Germany}

\author{Burkard~Hillebrands}
\affiliation{Fachbereich Physik and Forschungszentrum OPTIMAS, Technische Universit\"at Kaiserslautern, D-67663 Kaiserslautern, Germany}


\date{\today}

\begin{abstract}
We investigate the spin-wave excitation in microscopic waveguides fabricated by localized Cr$^\text{+}$ ion implantation in a ferromagnetic Ni$_\text{81}$Fe$_\text{19}$ film. We demonstrate that spin-wave waveguides can be conveniently made by this technique. The magnetic patterning technique yields an increased damping and a reduction in saturation magnetization in the implanted regions that can be extracted from Brillouin light scattering measurements of the spin-wave excitation spectra. Furthermore, the waveguide performance as well as the internal field of the waveguide depend on the doping fluence. The results prove that localized ion implantation is a powerful tool for the patterning of magnon spintronic devices. 
\end{abstract}

\pacs{}

\maketitle 

The field of spin dynamics and spintronics has greatly benefited from micro- and nanopatterning techniques. The advancement of research on small magnetic structures\cite{Mathieu1998, Demokritov2002, Demidov2010, Madami2011} strongly depends on the possibility to fabricate structures of accurate sizes and shapes. Based on these developments, the focus of many investigation has turned to the realization of spin logic devices, which use the electron spin\cite{Zutic2004} or spin waves as information carrier.\cite{Khitun2010, Serga2010, Lenk2011}

In order to establish a pure spin-wave logic architecture, there is a need for spin-wave conduits for the transport of spin information between the individual logic devices. Thus, the fabrication and investigation of spin-wave waveguides\cite{Vogt2009, Pirro2011, Sebastian2012, Clausen2011} has become a major research topic in the field of magnon spintronics. Currently, spin-wave waveguides with nanometer thicknesses and widths of several micrometers are fabricated by topographically patterning thin magnetic stripes onto a substrate that is usually nonmagnetic. However, the potential of other waveguide fabrication techniques have not been explored up to now, although they might contribute to an improvement of waveguide properties that were not accessible so far, such as, e.g., the inhomogeneity of the internal magnetic field.\cite{Guslienko2003} In that respect, ion implantation of ferromagnetic films is a promising technique for the fabrication of thin films with microscopic magnetic substructures.\cite{Fassbender2008, Neb2012, Folks2003} The different components contributing to the change in the magnetic properties due to Cr$^\text{+}$ ion implantation in Ni$_\text{81}$Fe$_\text{19}$ films have been thoroughly investigated.\cite{Fassbender2006}

\begin{figure}
	\includegraphics[viewport = 0 195 595 630, clip, scale=1.0, width=0.8\columnwidth]{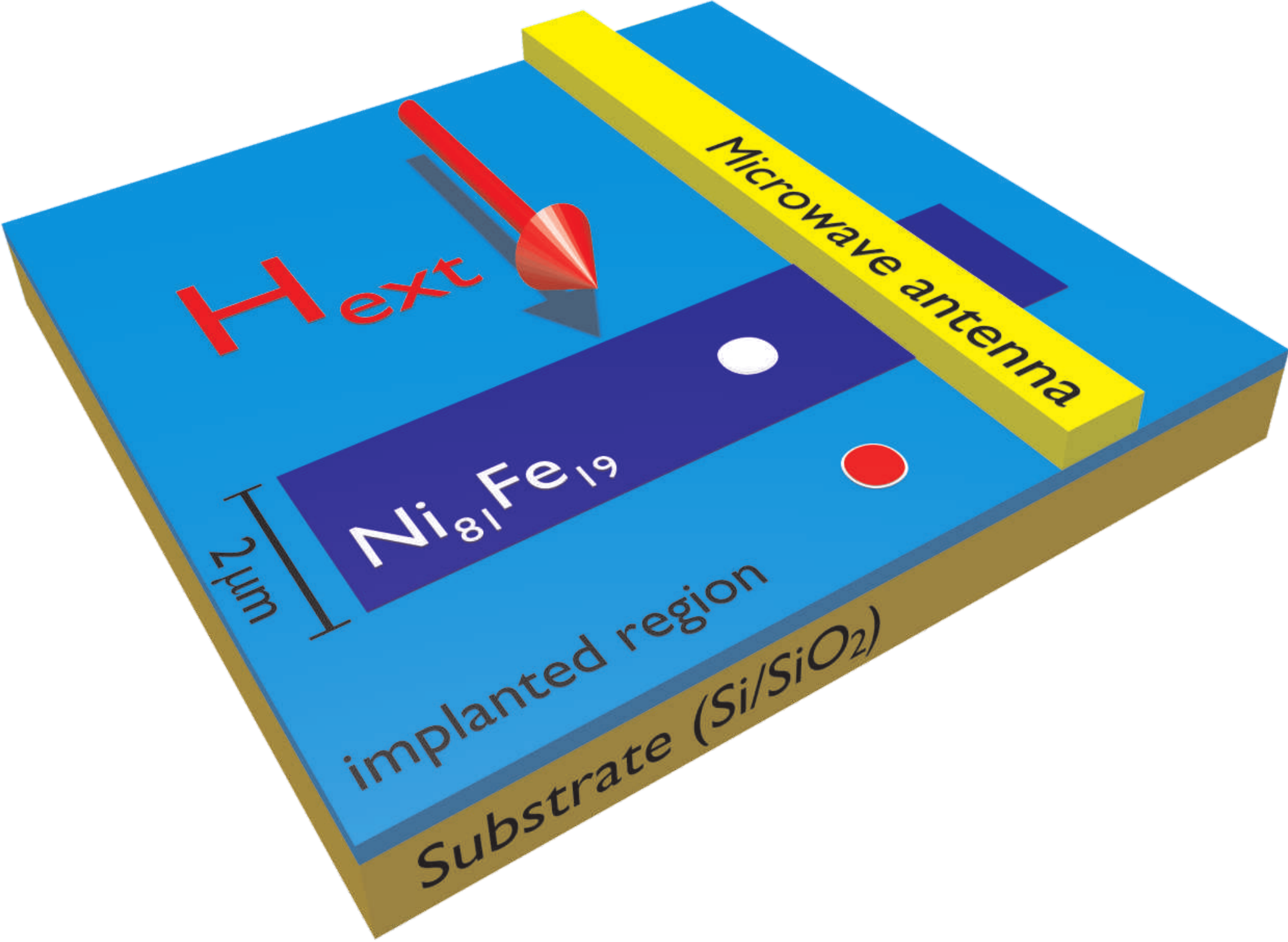}
	\caption{\label{Fig1} (Color online) Schematic sample setup. A Ni$_\text{81}$Fe$_\text{19}$ film with a thickness of 20\,nm is grown on top of a Si/SiO$_\text{2}$ substrate. By localized implantation of Cr$^\text{+}$ ions the magnetic properties of the implanted parts of the film are changed while a shielded region yields a 2\,$\mu$m wide Ni$_\text{81}$Fe$_\text{19}$ stripe with unchanged properties embedded in the film. A subsequently produced Cu microwave antenna serves as excitation source for spin waves. The two measurement positions for the investigations from Fig.\,\ref{Fig2} are marked by a red and white circle, respectively. The external magnetic field is applied along the short axis of the Ni$_\text{81}$Fe$_\text{19}$ stripe.
}
\end{figure}

In this letter a purely magnetic microscopic structuring of spin-wave waveguides by localized ion implantation in a Ni$_\text{81}$Fe$_\text{19}$ film is presented. We have designed a stripe shaped area with unchanged magnetic properties, embedded in a film with a reduced saturation magnetization $M_\text{S}$ and an increased damping $\alpha$. It is shown that the spin-wave distribution within the stripe has similar properties as in a topographically patterned Ni$_\text{81}$Fe$_\text{19}$ reference waveguide, if $M_\text{S}$ in the surrounding implanted region is decreased below at least 66\% of its original value. The spin-wave excitation spectra also indicate an influence of the reduced magnetization of the implanted region on the internal magnetic field of the Ni$_\text{81}$Fe$_\text{19}$ stripe. Hence, by localized ion implantation it is possible to create and adjust the properties of spin-wave waveguides by pure magnetic patterning.

A schematic of the sample setup is shown in Fig.\,\ref{Fig1}. A 20\,nm thick Ni$_\text{81}$Fe$_\text{19}$ film is grown onto a Si/SiO$_2$(200nm) substrate by molecular beam epitaxy (MBE). Subsequently, a lithographically structured polymethyl methacrylate (PMMA) resist mask is fabricated on top of the film via electron beam lithography. The magnetic patterning is achieved by exposing the sample to a broad beam irradiation of 30\,keV Cr$^\text{+}$ ions. This localized ion implantation in the areas which are not covered by the PMMA leads to the creation of $2\,\mu$m wide stripes with unchanged magnetic properties embedded in the implanted film. Due to the doping, this surrounding film has been modified in its saturation magnetization and damping. Using lift-off, a Cu antenna with a width of $1\,\mu$m and a thickness of $500$\,nm is fabricated across the stripes for the excitation of spin waves in the ferromagnetic film by the Oersted field of an applied microwave current. Depending on the ion fluence the modification of the Ni$_\text{81}$Fe$_\text{19}$ properties changes accordingly and, thus, the influence on the spin-wave properties varies. For this reason a series of samples has been fabricated with varying Cr$^\text{+}$ fluences of $2.2\cdot 10^{15}$\,ions/cm$^\text{2}$ (sample A), $6.6\cdot 10^{15}$\,ions/cm$^\text{2}$ (sample B) and $1.1\cdot 10^{16}$\,ions/cm$^\text{2}$ (sample C).

\begin{figure}
	\includegraphics[viewport = 116 20 472 826, clip, scale=1.0, width=0.8\columnwidth]{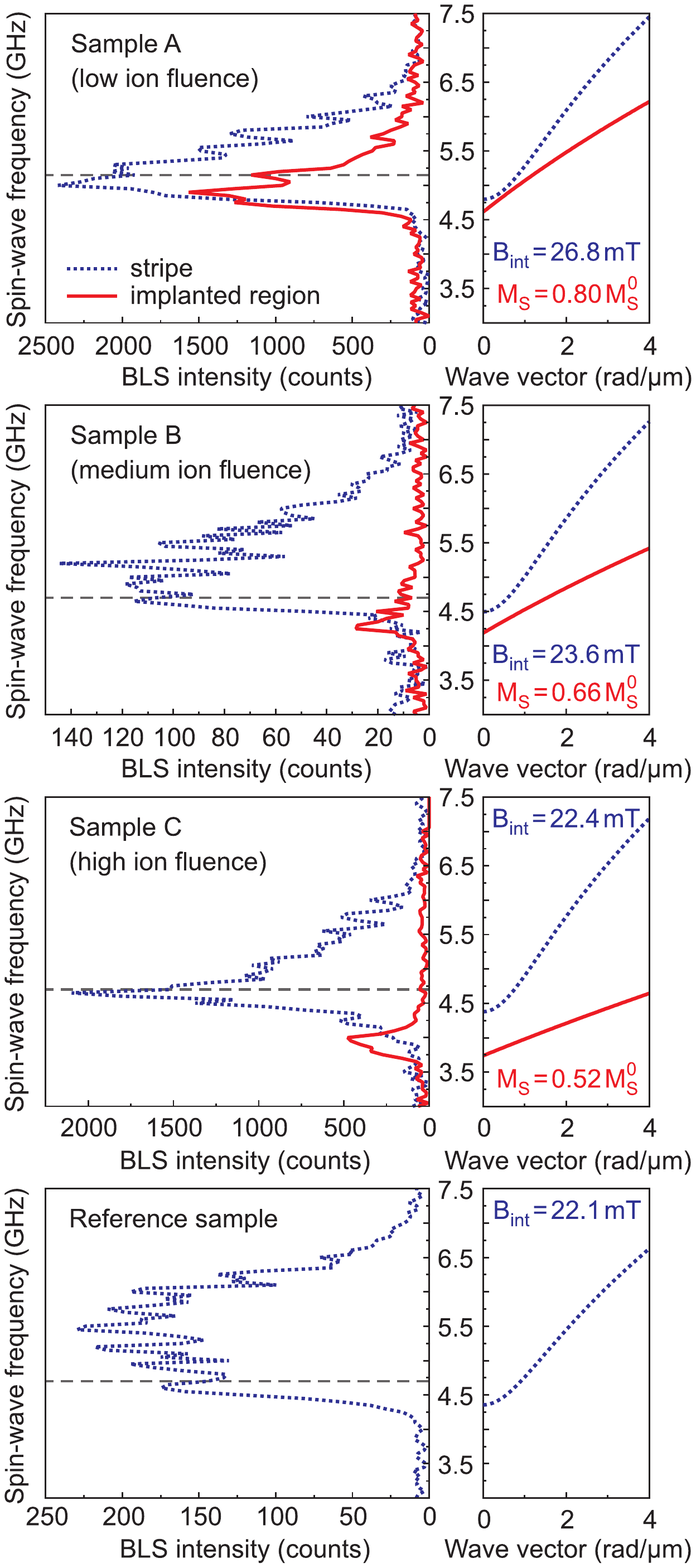}
	\caption{\label{Fig2} (Color online) Experimentally observed spin-wave resonance spectra (left column) and calculated dispersion relations (right column) of magnetically patterned samples with different ion fluences compared to a topographically patterned reference sample. The resonance spectra have been recorded on the non-irradiated stripe (dotted lines) as well as on the implanted region of the film (solid lines) using Brillouin light scattering microscopy. The dashed lines in the left column indicate the frequency of the measurements in Fig.\,\ref{Fig3}(a). The right column contains fitting values for the internal magnetic field and the normalized saturation magnetization.}
\end{figure}

For an initial characterization of the magnetization dynamics inside the modified film, the spin-wave excitation spectrum is recorded using Brillouin light scattering (BLS) microscopy. This technique is based on the inelastic scattering of laser light with spin waves and stands out due to its high sensitivity combined with a high spectral and spatial resolution.\cite{Demidov2004} The spin-wave excitation spectra are taken at the position of the stripe and in the implanted region, respectively, as indicated by the two circles in Fig.\,\ref{Fig1}. For this, an in-plane static magnetic field of $\mu_\text{0}H_\text{ext} = 30$\,mT is applied perpendicularly to the stripe's long axis, and a microwave current with varying frequency is applied to the antenna. The left column of Fig.\,\ref{Fig2} shows the resulting spin-wave excitation spectra. The doping causes a shift of the excitation spectra to lower frequencies in the irradiated areas. This allows for a quantitative analysis of the reduction of the saturation magnetization by calculating the spin-wave dispersion relation\cite{parameters, Kalinikos1986} and treating $M_\text{S}$ as a fit parameter to match the experimental data (see solid lines in the right column of Fig.\,\ref{Fig2}). With increasing ion fluence a reduction of $M_\text{S}$ to values of $0.80 M_\text{S}^\text{0}$ (sample A), $0.66 M_\text{S}^\text{0}$ (sample B) and $0.52 M_\text{S}^\text{0}$ (sample C) is determined, where $M_\text{S}^\text{0} = 860$\,kA/m is the saturation magnetization of the non-irradiated Ni$_\text{81}$Fe$_\text{19}$ film. Using the approach shown in Ref.\,\onlinecite{Marko2010}, the same magnetization values are found by polar magneto-optical Kerr effect measurements on equally implanted Ni$_\text{81}$Fe$_\text{19}$ films. Furthermore, the decrease in the BLS signal in the implanted region (solid lines) with respect to the unchanged stripe regions (dotted lines) indicates an increase in the damping constant $\alpha$ in accordance to Ref.\,\onlinecite{Fassbender2006}. However, a quantitative analysis of the change in $\alpha$ is not possible, since the absolute spin-wave intensities of different samples vary drastically and cannot be compared.

The results are in good agreement with previous investigations of Cr$^\text{+}$ implanted Ni$_\text{81}$Fe$_\text{19}$ films\cite{Fassbender2006} which, in addition, revealed a radiation-induced damage of the film surface and interface to a buffer layer. In the present case, atomic force microscopy (AFM) measurements of the surface profile show a thickness reduction of the Ni$_\text{81}$Fe$_\text{19}$ layer in the implanted area of up to 5\,nm. The reduced film thickness of the implanted region has been considered in the above calculations of the spin-wave dispersion relation. However, its influence is negligible compared to the change in $M_\text{S}$ for the dipolar dominated spin waves analyzed in this study. Moreover, a possible difference in the stripe width between different samples (caused by the varying irradiation fluence of the Cr$^\text{+}$ ions) can be excluded by the AFM measurements. 

Figure\,\ref{Fig2} also displays experimental spin-wave excitation spectra, which have been recorded on the non-irradiated stripe region (dotted lines) as well as on a topographically patterned Ni$_\text{81}$Fe$_\text{19}$ stripe serving as a reference. A comparison of the respective lower onset frequency (Ferromagnetic Resonance) of the spectrum shows that with increasing ion fluence the excitation spectrum is shifted to lower frequencies and approaches the spectrum of the reference sample. Since the stripe area has not been changed in its material properties, i.e. $M_\text{S}$ is unchanged, the influence on the spin-wave spectrum must be a result of the modification in the magnetic properties of the surrounding film. This long range interaction of the implanted film on the stripe region is reflected in the demagnetizing fields $B_\text{demag}$, which are created at the edges of the stripe. In the case of a small reduction of $M_\text{S}$ in the implanted regions the stray fields in the stripe will be negligible, resulting in an internal field $B_\text{int} = \mu_\text{0}H_\text{ext} + B_\text{demag} \approx \mu_\text{0}H_\text{ext}$. However, for a strong reduction in the saturation magnetization in the irradiated region a higher stray field originates from the boundaries of the stripe, which is directed opposite to the external field. Using $B_\text{int}$ as a fitting parameter in the calculation of the dispersion relation\cite{parameters2} allows for an evaluation of the internal field from the experimental data. The resulting $B_\text{int}$ values given in Fig.\,\ref{Fig2} confirm the above assumptions: With increasing ion fluence the internal magnetic field of the samples converges to the field value of the reference sample, which represents the minimum (i.e. zero) magnetic moment outside the stripe area and thus the highest possible stray-field induced reduction of $B_\text{int}$. Thus, the fabrication of spin-wave waveguides by ion implantation allows for a manipulation of the internal magnetic field inside the wave\-guide. 

\begin{figure}
	\includegraphics[viewport = 85 107 459 662, clip, scale=1.0, width=0.8\columnwidth]{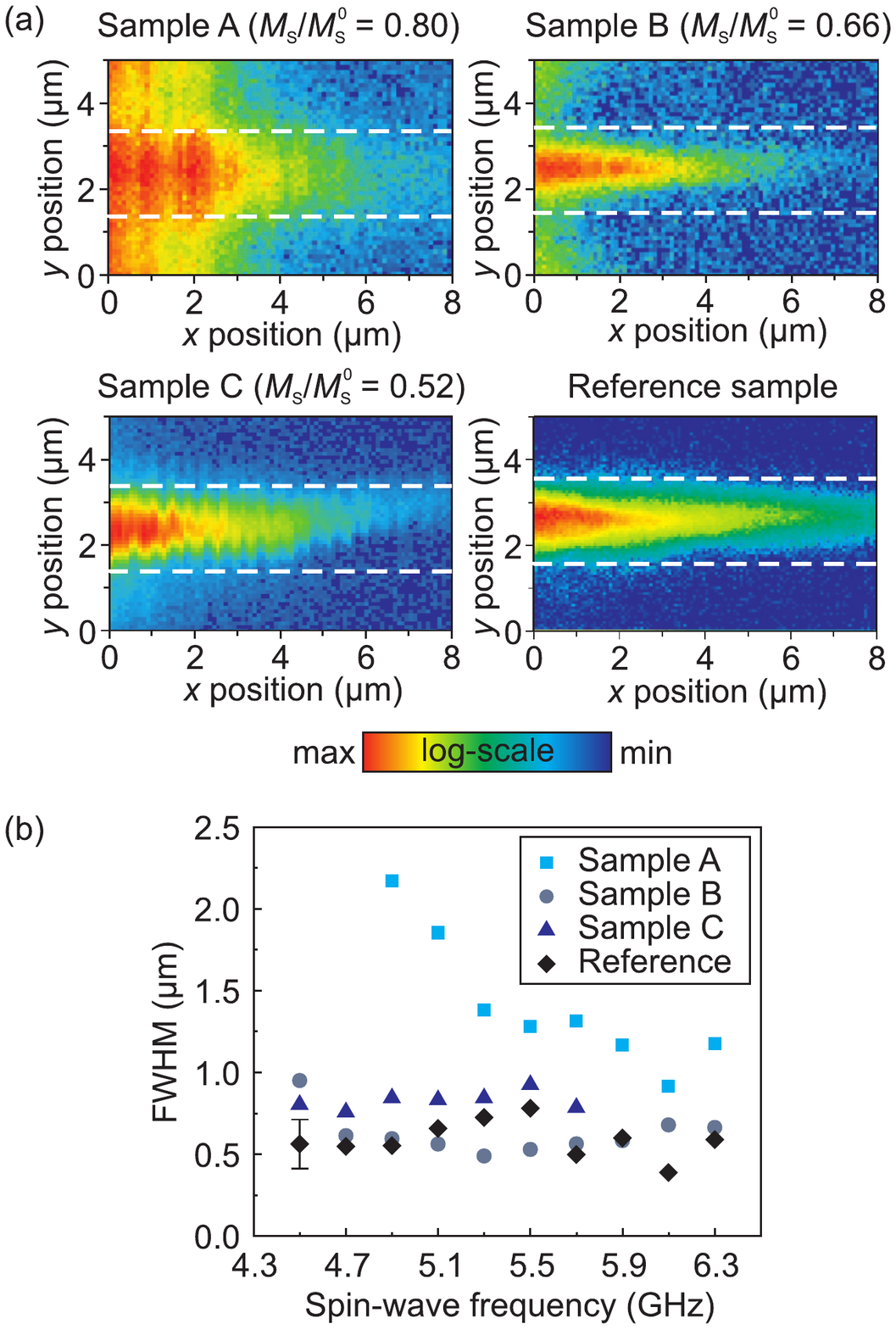}
	\caption{\label{Fig3} (Color online) (a) Color-coded spin-wave intensity distribution in the magnetically patterned Ni$_\text{81}$Fe$_\text{19}$ films and the reference waveguide. The position of the stripe with unchanged magnetic properties is indicated by the white dashed lines. Spin waves are excited by the antenna at $x = 0\,\mu$m. (b) Dependence of the width of the transverse spin-wave mode profiles on the spin-wave frequency. The width of the mode profiles has been measured by evaluating the full width at half maximum (FWHM). The experimental precision is indicated by the error bar.}
\end{figure}

As a next step, the spin-wave intensity distribution in the implanted films is investigated with respect to the suitability of the Ni$_\text{81}$Fe$_\text{19}$ stripe to act as a spin-wave waveguide. Spin waves are excited by the antenna with a frequency close to the Ferromagnetic Resonance, as indicated by the dashed lines in the left column of Fig.\,\ref{Fig2}. An in-plane static external magnetic field of $\mu_\text{0}H_\text{ext} = 30$\,mT is applied perpendicularly to the stripe's long axis (Damon-Eshbach geometry). The spin-wave propagation along the stripe is studied using BLS microscopy by recording two-dimensional maps of the spatial spin-wave intensity distribution with a resolution of about 300\,nm. Beside the implanted films the topographically patterned reference waveguide has also been studied. The results are shown in Fig.\,\ref{Fig3}(a). The spin-wave intensity is color coded and plotted as a function of the position on the film with the antenna being at $x = 0$\,mm, the dashed lines indicate the stripe position. Compared to the reference waveguide, the spin-wave intensity pattern in the case of sample A differs significantly. Although a higher decay length can be assumed in the stripe region, there is significant excitation of spin waves in the implanted regions. In contrast, the spin-wave intensity distribution is concentrated to the stripe region in the case of samples B and C and resembles the reference waveguide pattern, indicating a waveguide-like behavior. This change in the waveguide performance can be understood using the results of Fig.\,\ref{Fig2}. For each sample the respective excitation frequency is indicated by a dashed line. While for sample A the excitation efficiency in both the implanted and the stripe region is similar, a pronounced difference is obtained in the case of samples B and C. This is attributed to both a change in $M_\text{S}$ and in $\alpha$, since the former prevents the spreading of spin waves in the regions outside the stripe by shifting and narrowing the excitation spectrum in frequency while the latter decreases their lifetime. Consequently, the excitation and subsequent propagation of spin waves is confined to the stripe regions, leading to a waveguide-like behavior.

For a more quantitative view on the waveguide performance, the transverse spin-wave intensity profiles are measured by scanning the BLS laser focus along the short axis of the stripe. The width of the intensity distribution is analyzed by determining the full width at half maximum (FWHM). Figure\,\ref{Fig3}(b) depicts the resulting FWHM values as a function of the applied spin-wave excitation frequency. In the case of sample A the FWHM strongly decreases with increasing excitation frequency. This deviates from the observation on samples B and C, where the width of the profiles is unaffected by the change in the microwave frequency. The latter also exhibit the same FWHM values as the reference waveguide within the experimental precision as indicated by the error bar. In the reference waveguide, the lateral confinement of the spin waves along the waveguide width leads to the formation of quantized transverse waveguide modes.\cite{Mathieu1998} The width of these modes depends on the transverse wave vector component, which is quantized along the short stripe axis and hence constant for a given mode number. Obviously, this quantization along the short stripe axis is given in samples B and C due to the mismatch of the dispersion relations of the implanted region and that of the stripe, respectively, (see Fig.\,\ref{Fig2}), which prevents the propagation of spin waves excited in the stripe into the implanted regions. However, in the case of sample A, the two dispersion relations converge with decreasing frequency, resulting in an increased spreading of spin waves over the whole film in accordance to the experimental results shown in Fig.\,\ref{Fig3}(b). Thus, the results prove that the stripes of samples B and C can be considered as genuine spin-wave waveguides and that a reduction of the saturation magnetization in the implanted area to $66\%$ of its original value is sufficient to channel the spin waves inside the magnetically patterned Ni$_\text{81}$Fe$_\text{19}$ stripe.

In conclusion, a purely magnetic patterning of a spin-wave waveguide has been achieved by localized Cr$^\text{+}$ ion implantation in a Ni$_\text{81}$Fe$_\text{19}$ film. The waveguide performance depends on the induced changes in the magnetic properties and hence can be controlled by the doping intensity. Moreover, having a waveguide inside a region with a nonzero saturation magnetization allows for an adjustment of the internal field of the waveguide. These results show that the fabrication of spin-wave waveguides by means of localized ion implantation provides potential for the manipulation of various waveguide properties and thus pioneers applications for spin-wave conduits in the field of magnon spintronics.



%
%

%


The authors wish to thank the Nano Structuring Center (NSC) of the TU Kaiserslautern for support with the sample preparation and R. Neb for the evaporation of the Ni$_\text{81}$Fe$_\text{19}$ films. Financial support by the DFG within the Graduiertenkolleg (GRK~792) is gratefully acknowledged.



\end{document}